\documentstyle[aps]{revtex}

\newcommand{\PC}[1]{$\footnotemark\footnotetext{PC: #1}$}

\newcommand{\tr}{{\rm tr}\, }

\newcommand{\Tr}{\mbox{Tr}\,}
\newcommand{\Id}{\frac{dw}{2 \pi i}}
\newcommand{\Lop}{{\cal L}}
\newcommand{\sign}[1]{{\rm sign}({#1})}
\newcommand{\dzeta}{dynamical zeta function}
\newcommand{\qS}{Gutzwiller-Voros zeta function}
\newcommand{\Gt}{Gutzwiller trace formula}
\newcommand{\Zqm}{Z_{qm}}
\newcommand{\Vd}{Vattay determinant}
\newcommand{\Fd}{Fredholm determinant}
\newcommand{\FD}{FREDHOLM DETERMINANT}
\newcommand{\Fqm}{F_{qm}}
\newcommand{\qFd}{quantum Fredholm determinant}

\newcommand{\cFd}{classical Fredholm determinant}
\newcommand{\zfct}[1]{\zeta ^{-1}_{#1}}
\newcommand{\ExpaEig}{\Lambda}

\newcommand{\rf}[1]{~\cite{#1}}

\newcommand{\barr}{\begin{array}}
\newcommand{\ear}{\end{array}}
\newcommand{\bea}{\begin{eqnarray}}
\newcommand{\nnu}{\nonumber}
\newcommand{\eea}{\end{eqnarray}}
\newcommand{\beq}{\begin{equation}}
\newcommand{\continue}{\nonumber \\ }

\newcommand{\eeq}{\end{equation}}
\newcommand{\ee}[1] {\label{#1} \end{equation}}
\newcommand{\refeq}[1]{(\ref{#1})}

\newcommand{\combinatorial}[2]{
   \left(
   \begin{array}{c}
      {#1}  \\ {#2}
   \end{array}
   \right) }

\begin{document}
\twocolumn[

\title{    A Fredholm Determinant for Semi-classical Quantization}

\author{     Predrag Cvitanovi\'c,
             Per E. Rosenqvist,
             G\'abor Vattay
       }
\address{       Niels Bohr Institute,
                Blegdamsvej 17, DK-2100 Copenhagen \O, Denmark }

\author{      Hans Henrik Rugh}
\address{       Unit\'e de Math\'ematiques Pures et Appliqu\'ees,
                Ecole Normale Sup\'erieure de Lyon\\
                46, All\'ee d'Italie,
                F-69364 Lyon, France }
\date{\today}
\maketitle
\begin{abstract}
We investigate a new type of approximation to quantum determinants,
the ``\qFd", and test numerically
the conjecture that for Axiom A hyperbolic flows
such determinants have a larger domain of analyticity and
better convergence than
the \qS s derived from the \Gt.
The conjecture
is supported by numerical investigations of the 3-disk repeller,
a normal-form model of a flow, and a model 2-$d$ map.
\end{abstract}
]

\narrowtext
\section{INTRODUCTION}

\PC{version 16 july 1993, ~file chaos.nbi.dk:\\
/users/predrag/tex/articles/q\_fredholm/qfred6.tex
        }
While various periodic orbit formulas may be formally equivalent,
in practice some are vastly preferable to others.
Trace formulas, such as the thermodynamic averages in
classical dynamics, and the semi-classical \Gt\
in quantum mechanics are difficult to use for anything
other than the leading eigenvalue estimates.
However, \dzeta s\rf{ruelle}, \Fd s\rf{grot}
and the \qS \rf{gutbook,voros} have recently been
established as powerful tools for evaluation of classical and
quantum averages in low dimensional chaotic dynamical
systems\rf{AACI}~-\rf{CHAOS92}.
Recent advances include
new, cycle expansion\rf{cycprl} based numerical spectra evaluations and
Riemann conjecture inspired functional equations.
The convergence of cycle expansions
of zeta functions and
\Fd s depends on their analytic properties; particularly strong
results exist for Axiom A hyperbolic systems, for which the
\dzeta s are meromorphic\rf{Ruelle76},
and the \cFd s
are entire functions\rf{frie,Rugh92}. No such results exist for
the determinants used in quantum theory, but formal analogies
to the classical case have led to introduction
of the {\em \qFd}\rf{CR92}
\beq
\Fqm(E)  =  \prod_p \prod_{k=0}^{\infty}
           \left(1 - {
                    e^{ -\frac{i}{\hbar} S_p(E) + i \pi m_p/2}
              \over |\Lambda_p|^{1/2} \Lambda_p^k }
           \right)^{k+1}
\ee{Fred_qm}
as an alternative to the \qS
\beq
\Zqm(E)  =  \prod_p \prod_{k=0}^{\infty}
           \left(1 - {
                    e^{ -\frac{i}{\hbar} S_p(E) + i \pi m_p/2}
              \over |\Lambda_p|^{1/2} \Lambda_p^k }
           \right)
\,.
\ee{Selb_qm}
We present here the numerical evidence in support of the conjecture
of ref.~\cite{CR92}:\\

\noindent
{\em
For Axiom A systems the
\qFd\
has a larger domain of analyticity than the \qS.}
\\

We shall consider here only purely hyperbolic flows with
the topology of a Smale horseshoe.
The important conceptual insight of Smale\cite{smale} is the
realization that for such flows the associated zeta
functions have nice analytic structure. In a more
formal setting, such flows are called ``Axiom A", and
Ruelle\cite{ruelle} proves that for expanding analytic maps
the zeta functions are meromorphic, and the spectrum is discrete.
This differs very much from the intuition acquired in studies
of quantum chaos; there is no
``abscissa of absolute convergence" and no ``entropy wall",
the exponential proliferation of orbits can be controlled,
and the Selberg-type zeta
functions are entire and converge everywhere.

In classical mechanics and number theory, the zeta functions
and the \Fd s are exact. All quantum mechanical studies take a
saddle-point approximation - the \Gt\  -
as the starting point, and for quantum mechanics
an important conceptual problem arises already at
the level of derivation of the \qS s; how accurate are they,
and can the periodic orbit theory of such semi-classical approximations
be systematically improved? We shall
not address this problem here; in this paper we are interested in the
convergence of cycle expansion truncations of
the \qS.
The problem is {\em classical} in the
sense that all quantities used in periodic orbit calculations - actions,
stabilities, Maslov phases - are classical quantities.

The main limitation of the study presented here is that
the \Fd s are proven to be entire only for hyperbolic flows
with symbolic dynamics with finite grammar (Axiom A flows). Hence
our numerics is restricted to three systems
with complete binary grammar and bounded nonlinearity:
\\

(a) the 3-disk repeller

(b) a Hamiltonian H\'enon map as a normal form approximation to a flow

(c) a model 2-$d$ hyperbolic mapping
\\

The paper is organized as follows:
in sect.~\ref{FLOWS} we review the evolution
operator formalism for smooth flows.
In sect.~\ref{ENTIRE} we explain the theorems that guarantee
that \Fd s for Axiom A systems are entire.
In sect.~\ref{DETERMINANTS} we define and motivate the
determinants used in this paper.
In sect.~\ref{CYCLE} we review the cycle expansions, and give
convergence estimates for various expansions.
In sect.~\ref{NUMERICAL} we discuss the numerical evidence in support
of the \qFd\ conjecture.

\section{FLOWS, EVOLUTION OPERATORS AND THEIR SPECTRA}
\label{FLOWS}

Functional determinants and zeta functions arise in
classical and quantum mechanics because in both the dynamical evolution
can be described by action of linear evolution operators
on infinite-dimensional vector spaces.
The classical {\em evolution operator} for a $d$-dimensional map or
a $(d+1)$-dimensional flow is given by:
\beq
\Lop^t (y,x) =  \delta ( y - f^t(x)) g^t(x)
\,\, .
\ee{TransOp}
For discrete time, $f^n(x)$ is $n$-th iterate of the map $f$; for
continuous flows, $f^t(x)$ is the trajectory of the initial point $x$.
$g^t(x)$ is a weight multiplicative along the trajectory; its
precise functional form depends on the dynamical average
under study. For purposes of this section it suffices to
take $g^t(x)=1$, essentially the Perron-Frobenius operator case.

The global averages (escape rates, energy eigenvalues, resonances,
fractal dimensions, etc.) can be extracted from the eigenvalues of the
evolution
operators. The eigenvalues are
given by the zeros of appropriate determinants. One way to evaluate
determinants is to expand them in terms of traces, {\em log det = tr log},
and in this way the spectrum of an evolution operator becomes related to
its traces, {\em ie}. periodic orbits.
Formally, the traces tr$\Lop^t$ are easily evaluated as integrals
of Dirac delta functions as follows:

\subsection{Trace formula for maps}

If the evolution is given by a discrete time mapping, and
all periodic points are known to have stability eigenvalues
$\Lambda_k \neq 1$
strictly bounded away from unity,
the trace $\Lop^n$ is given by
the sum over all periodic points $x$ of period $n$:
\bea
\tr \Lop^n &=& \int dx dy \delta (x-y) \Lop^n (y,x)
                \continue
             &=&  \sum_p n_p \sum_{r=1}^\infty
        {\delta_{n,n_p r}
           \over |\det \left( {\bf 1}-{\bf J}_p^{r} \right)| }
\,\, ,
\label{tr_L}
\eea
where
\beq
    {\bf J}_p(x)= \prod_{j=0}^{n_p-1}  {\bf J}(f^{j}(x)), \quad
    J_{kl}(x) = {{\partial } \over {\partial x_l}} f_k(x)
\label{jacob}
\eeq
is the [$d$$\times$$d$] Jacobian matrix evaluated at the
periodic point $x$,
and the product goes over all periodic points $x_i$ belonging to a
given prime cycle $p$.
The {\em trace formula} is the Laplace transform of $\tr \Lop^t$
which, for discrete flows, is simply the generating function
\[
\tr \Lop(z) = \sum_{n=1}^\infty z^n \tr \Lop^n
        = \sum_{\alpha=0}^\infty {z e^{-\nu_\alpha} \over 1 - z e^{-\nu_\alpha}
}
\]
where $e^{-\nu_0}$, $e^{-\nu_1}$, $e^{-\nu_2}, \dots$ are the
eigenvalues of $\Lop$.  For large times
$\, \det \left( {\bf 1}-{\bf J}^{(n)}(x_i) \right)
        \rightarrow \Lambda_i $, where $\Lambda_i$ is the product
of the expanding eigenvalues of ${\bf J}^{(n)}(x_i)$, so the
trace is dominated by
\beq
\tr \Lop(z) \approx
        \sum_{n=1}^\infty z^n \sum_{x_i \in \mbox{\small Fix}(f^n)}
                                   {1 \over |\Lambda_i|}
        = {z e^{-\nu_0} \over 1 - z e^{-\nu_0} } + \, \dots
\,\,,
\ee{thermo}
and diverges at the leading eigenvalue $1/z =e^{-\nu_0} $. This
approximation, which in current physics literature is called
the ``thermodynamic" 
or the ``$f$ of $\alpha$" formalism\rf{chicago5}, is adequate (but far
from optimal) for extraction of the leading eigenvalue of $\Lop$,
and difficult to apply to extraction of the non-leading eigenvalues.

\subsection{Trace formula for flows}

For flows the eigenvalue corresponding to the eigenvector
along the flow (the velocity vector) necessarily equals
unity for all periodic orbits, and therefore
the integral  \refeq{tr_L} requires a more careful
treatment\rf{CEflows}.

To evaluate the contribution of a prime
periodic orbit $p$ of period $T_p$,
one choses a local coordinate system with a longitudinal coordinate
$dx_\parallel$ along the direction of the flow,
and $d$ transverse coordinates ${x}_\perp$
\beq
\tr_p \Lop^{t}
= \int_{V_p} d{x}_\perp dx_\parallel
\delta({x}_\perp -{f}_\perp ^{t}({x}))
\delta({x}_\parallel-{f}_\parallel^{t}({x}))
\,\, .
\label{KT4}
\eeq
Integration is restricted to an infinitesimally thin tube $V_p$
enveloping the cycle $p$.

Let  $v = |{\bf F}({x})|$ be the velocity along the orbit, and change
the longitudinal variable to $dx_\parallel
= v d\tau$. Whenever the time
$t$ is a multiple of the cycle period $T_p$,
the integral along the trajectory yields
\bea
\int_{V_p} dx_\parallel \delta({x}_\parallel -{f}_\parallel^t({x}))
 &=& \sum_{r=1}^\infty \delta (t-r T_p)  \int_p d\tau
        \continue
 &=& T_p \sum_{r=1}^\infty \delta (t-r T_p)
\,\, .
\label{long}
\eea

Linearization of the
periodic flow in a plane perpendicular to the orbit yields
the same weight as
for the maps:
\beq
\int_{V_p} d{x}_\perp \delta({x}_\perp -{f}_\perp ^{-r T_p} ({x}))
  =   {1 \over {|\det \left( {\bf 1}-{\bf J}^r_p \right)| }}
\,\, ,
\label{trans}
\eeq
where ${\bf J}_p$ is the $p$-cycle [$d$$\times$$d$] transverse Jacobian,
and we have assumed hyperbolicity, {\em ie}. that
all transverse eigenvalues are bounded away from unity.
A geometrical interpretation of weights such as (\ref{trans}) is that after the
$r$-th return to a surface of section, the initial tube $V_p$ has been
stretched out along the expanding eigendirections,
with the overlap with the initial volume given by
$1 / |\det \left( {\bf 1}-{\bf J}^{r}_p \right)| $.

Substituting (\ref{long}-\ref{trans})
into (\ref{KT4}), we obtain an expression for
${\rm tr\,}{\cal L}^{t} $ as a sum over all prime cycles $p$
and their repetitions 
\[
\tr \Lop^{t}
     =   \sum_{p} T_p \sum_{r=1}^\infty {
               \delta (t-r T_p)
         \over {|\det \left( {\bf 1}-{\bf J}_p^{r} \right)| }}
\, \, .
\]
A Laplace transform replaces the above sum of Dirac delta functions
by the {\em trace formula for classical flows}\rf{CEflows}:
\beq
\tr \Lop(s)
        = \int_0^\infty dt e^{st} \tr \Lop^t
             = \sum_p T_p \sum_r^{\infty}
        {e^{sT_pr} \over |\det \left( {\bf 1}-{\bf J}_p^r \right)| }
\,\, .
\ee{tr_L_cont}
We should caution the reader that in taking the Laplace transform
we have ignored a possible
$
{t \rightarrow 0_{+}} $
volume term,
as we do not know how to regularize the delta function kernel
in this limit.  In the quantum (or heat
kernel) case this limit gives rise to the Weyl or Thomas-Fermi mean
eigenvalue spacing.
A more careful treatment might assign such volume term some interesting role
in the theory of classical resonance spectra.

The semi-classical evaluation of the quantum trace is considerably
more laborious, but the final result,
given in sect.~\ref{QUANTUM},
is very similar in form to the above classical trace.

\subsection{\Fd s}

The problem with the classical
\refeq{thermo}, \refeq{tr_L_cont}
and the \Gt s \refeq{tr_Gutz}
is that
they diverge precisely where one would like to use them
(we return to this in sect.~\ref{ABSCISSA}). While
in the physics literature on dynamically generated strange sets
this does not prevent numerical extraction of reasonable
``thermodynamic" averages, in
the case of the \Gt\ this leads to the perplexing
observation that crude estimates of the radius of convergence
seem to put the entire physical spectrum out of reach. This
problem is cured by going from trace formulas to determinants,
which turn out to have larger analyticity domains. For maps, the two are
related by
\[
F(z) = \det(1-z\Lop) =
  \exp\left(-\sum_n^\infty {z^n \over n} \tr \Lop^n \right)
\]
For flows the \cFd\ is given by
\beq
F(s) = {\rm exp}  \left( - {
         \sum_{p} \sum_{r=1}^\infty {1 \over r}
 {  e^{ s T_p r }   
 \over  { | \det \left( {\bf 1}-{\bf J}_p^{r} \right) | } }
         } \right)
\,\,  ,
\label{Z(s)}
\eeq
and the classical trace formula \refeq{tr_L_cont}
is the logarithmic derivative of the \cFd
\beq
\tr \Lop(s) = {d \over ds} \log F(s)
\,\, .
\ee{der_det}
With $z$ set to $z=e^s$, the \Fd\ \refeq{Z(s)} applies both to maps and flows.
A \Fd\ can be rewritten as an infinite product over periodic orbits, by noting
that the $r$ sum in  (\ref{Z(s)}) is close in form to expansion of a logarithm.
We cast it into such form by expanding the Jacobian weights
in terms of stability eigenvalues.
For a 3-dimensional Hamiltonian flow with one expanding eigenvalue
$\Lambda$,
and one contracting eigenvalue $1/\Lambda$,
the weight in (\ref{Z(s)}) may be expanded as follows:
\bea
{1 \over |\det \left( {\bf 1}-{\bf J}_p^{r} \right)|}
 &=&  {1 \over |\Lambda|^r (1-1/\Lambda_p^{r})^2 }
                \continue
 &=&   {1 \over |\Lambda|^r} \sum_{k=0}^\infty (k+1) \Lambda_p^{-kr}
\,\, .
\label{Jac_exp}
\eea
With this we can rewrite the \Fd\ exponent as
\[
     \sum_{r=1}^\infty {1 \over r}
        {       e^{s T_p r}
         \over |\det \left( {\bf 1}-{\bf J}_p^{r} \right)| }
 =  
     \sum_{k=0}^\infty (k+1)
         \log \left( 1 -  { e^{ s T_p } 
                            \over
                           |\Lambda_p| \Lambda_p^{k} } \right)
\,\, .
\]
and represent the \Fd\ as a Selberg-type product\rf{CEflows}
\bea
F(s) &=&  \prod_p  \prod_{k=0}^\infty
                        \left( 1 -   t_p /\Lambda_p^k \right)^{k+1}
\, ,
        \continue
t_p &=& {e^{s T_p} \over |\Lambda_p| } z^{n_p}
\,\,.
\label{cl_weig}
\eea
$z$ is a book-keeping variable that we will use to
expand zeta functions and determinants, set to $z=1$
in calculations. In general, $t_p$ depends on the dynamical average
one wishes to evaluate; this particular weight is used in evaluation
of escape rates and correlation spectra\rf{CPR90,eck}.
\\

The above heuristic manipulations are potentially dangerous, as
we are dealing with infinite-dimensional vector spaces and
singular integral kernels;
we outline now
the ingredients of the
proofs that put the above formulas on solid mathematical footing.

\section{\FD S CAN BE ENTIRE}
\label{ENTIRE}

As the introduction of the \qFd\ \refeq{Fred_qm}
is motivated by its close analogy with the
\cFd s, we shall sketch here the basic ideas behind the
proofs that the \cFd s are entire,
without burdening the reader with too many technical details
(rigorous treatment is given in refs.~\cite{Ruelle76,frie,Rugh92}).
The reason why one cares whether a \Fd\ is entire or not is
that in practice one can extract many more eigenvalues,
and to a higher accuracy, from
entire \Fd s than from functions which are not entire.
The main point of the theorems explained below
is that the \Fd s are entire
functions in any dimension, provided that
\\

\noindent
1. the evolution operator is {\em multiplicative} along the flow,
\\
\noindent
2. the symbolic dynamics is a {\em finite subshift},
\\
3. all cycle eigenvalues are {\em hyperbolic} (sufficiently bo\-und\-ed
away from 1),
\\
4. the map (or the flow) is {\em real analytic}, {\em ie.}
   it has a piecewise  analytic continuation to a complex extension
   of the phase space.
\\

The notion of Axiom A systems is a mathematical abstraction
of 2 and 3. It would take us too far to give and explain
the definition of the Axiom A systems
(see refs.~\cite{smale,bowen}). Axiom A implies, however,
the existence of a Markov partition of the phase space
(see below) from which 2 and 3 follow. Properties 1 and 2
enable us to represent the evolution operator as a matrix
in an appropriate basis space; properties 3 and 4 enable us
to bound the size of the matrix elements and control the
eigenvalues.
To see what can go wrong consider the following examples:

Property 1 is violated for flows in 3 or more dimensions by
the following weighted evolution operator
\beq
\Lop^t (y,x) =  |\Lambda^t(x)|^{\beta} \delta ( y - f^t(x))
\,\, ,
\label{bad_weight}
\eeq
where $\Lambda^t(x)$ is an eigenvalue of the Jacobian transverse
to the flow.
While for the Jacobians $J_{ab}=J_a J_b$ for
two successive segments $a$ and $b$ along the trajectory, the
corresponding eigenvalues are in general {\em not} multiplicative,
$\Lambda_{ab} \neq \Lambda_a \Lambda_b$ (unless $a$, $b$ are repeats
of the same cycle, so $J_a J_b = J^{n_a+n_b}$),
so the above evolution operator is not multiplicative along the trajectory.
The theorems require
that the evolution be represented as a matrix in appropriate polynomial
basis, and thus cannot be applied to non-multiplicative kernels,
{\em ie}. kernels that do not satisfy the semi-group property
$\Lop^{t'} \Lop^t = \Lop^{t'+t}$.

Property 2 is violated by the 1-$d$ map
\[ f(x) = \alpha (1 \; - \; |1-2x|) \ , \ \ \ 1/2 < \alpha < 1 \ .\]
All cycle eigenvalues are hyperbolic, but
the critical point $x_c=\frac{1}{2}$ is in general not
a pre-periodic point, there is no finite Markov partition,
the symbolic dynamics does not have a finite grammar, and the
theorems discussed below do not apply. In practice this means that
while the leading eigenvalue of $\Lop$ might be computable,
the reminder of the spectrum is very hard to control; as the
parameter $a$ is varied, nonleading zeros of the Fredholm
determinant move wildly about.

Property 3 is violated by the map
\[
   f(x) = \left\{ \begin{array}{ll}
                    x + 2 x^2 & \ , \ \  \ x \in I_0 = [0,\frac{1}{2}]\\
                    2 - 2 x   & \ , \ \ \  x \in I_1 = [\frac{1}{2},1]
                 \end{array} \right. .
\]
Here the interval $[0,1]$ has a Markov partition into the
two subintervals $I_0$ and $I_1$ on which $f$ is monotone.
However, the fixed point at $x=0$ has stability $\Lambda=1$,
and violates the condition 3. This type of map is called intermittent
and necessitates much extra work\rf{PS92}. The problem is that
the dynamics in the neighborhood of a marginal fixed
point is very slow, with correlations decaying as power laws
rather than exponentially.

The property 4 is required as
from a mathematical point of view,
the heuristic approach of sect.~\ref{FLOWS} faces two major hurdles:
\begin{enumerate}
\item
   The trace \refeq{tr_L} is not well defined since  the integral kernel
   is singular.
\item
   The existence and properties of eigenvalues are by no means clear.
\end{enumerate}

Both problems are related to how one defines the function space on
which the evolution operator acts.
As in physical applications one studies smooth dynamical observables,
we restrict the space to smooth functions,
more precisely, the space of functions analytic
in a given complex domain and having a continuous extension
to the boundary of the domain.
In practice ``real analytic'' means that
all expansions are polynomial expansions.
In order to illustrate how this works in practice, we first work out
a few simple examples.

\subsection{Expanding maps}

We start with the trivial
example of a repeller with only one expanding linear branch
\[  f(x) = \ExpaEig x \ \ \ \ |\ExpaEig| > 1 \ .\]
The action of the associated evolution operator is
\[
\Lop \phi (y) =  \int dx \delta(y-\Lambda x)\phi(x)
              = {1 \over |\ExpaEig|} \phi (y/\ExpaEig) \,.
\]
{}From this one immediately identifies eigenfunctions and eigenvalues:
\beq
 \Lop \, y^n = {1 \over |\ExpaEig| \ExpaEig^{n}} y^n  \,,
\ \ \ \ n = 0,1,2,\ldots
\ee{FP_eigs}
The ergodic theory, as presented by Sinai\rf{sinai} and others,
tempts one to use a space of either
integrable or square integrable functions. For our
purposes, this space is too big;
had we not insisted on analyticity, non-integer and even complex
powers could be used in the construction. In particular,
in the space $L^1$ all $|\Lambda|^{\alpha-1}$ with $\alpha$ complex
but $\mbox{Re}(\alpha) < 1$ would be eigenvalues, {\em ie.}
the eigenvalues would fill out the unit disk.
We note that the eigenvalues $\ExpaEig^{-n-1}$
fall off exponentially with $n$, and
that the trace of $\Lop$ is given by
\[ \tr \Lop = {1 \over |\ExpaEig|} \sum_{n=0}^\infty \ExpaEig^{-n}
                  = {1 \over |\ExpaEig| (1-\ExpaEig^{-1})}
                   = {1 \over |f' - 1|} \]
in agreement with \refeq{tr_L}. A similar result is easily obtained
for powers of $\Lop$, and for the \Fd\ one obtains:
\[
 \det(1-z\Lop) = \prod_{k = 0}^\infty (1-{z\over|\ExpaEig| \ExpaEig^{k}}) =
        \sum_{k = 0}^\infty c_k t^k
\,,
\]
$ t =-z/|\ExpaEig|$,
where the coefficients $c_k$ are given explicitly by the Euler
formula\rf{eule}
\beq
 c_k = \frac{        1      }{1-\ExpaEig^{-1}} \;
         \frac{\ExpaEig^{-1}}{1-\ExpaEig^{-2}} \; \cdots \;
         \frac{\ExpaEig^{-k+1}}{1-\ExpaEig^{-k}}  \ \ \ .
\ee{Euler}
The coefficients decay asymptotically {\em faster} than exponentially,
as $\ExpaEig^{-k(k-1)/2}$.  This property ensures that for a
repeller consisting of a single repelling point the \cFd\
is entire in the complex $z$ plane.

While it is not at all obvious that what is true for a single
fixed point should also apply to a Cantor set of periodic points,
the same asymptotic decay of expansion coefficients
is obtained when several expanding branches are involved.
Consider a monotone and expanding 1-$d$ map $f(x)$, with
$|f'_\epsilon(x)|>1$  on two non-overlapping intervals
\beq
f(x) =  \left\{ {
        f_0(x) , \quad  x \in I_0
        \atop
        f_1(x) , \quad  x \in I_1
        } \right.
\, .
\ee{1d_rep}
The simplest non--trivial example is a piecewise-\-linear
2--branch repeller with slopes $\Lambda_0$ and $\Lambda_1$.
By the chain rule $\Lambda_p= \Lambda_0^{n_0} \Lambda_1^{n_1}$,
where the cycle $p$ contains $n_0$ symbols $0$ and $n_1$ symbols $1$,
so the trace \refeq{tr_L} reduces to
\bea
\tr \Lop^n &=&  \sum_{m=0}^{n} \combinatorial{m}{n}
        {1 \over | 1-\Lambda_0^{m} \Lambda_1^{n-m}| }
                \continue
             &=&  \sum_{k=0}^{\infty} \left(
        {1 \over |\Lambda_0| \Lambda_0^{k}} +
        {1 \over |\Lambda_1| \Lambda_1^{k}}
                                      \right)^n
\,.
\nnu
\eea
The \Fd\ \refeq{Z(s)} is given by
\[
\det(1-z\Lop) = \prod_{k=0}^\infty
\left( 1 - {t_0 \over \Lambda_0^k} - {t_1 \over \Lambda_1^k} \right)
\,\,,
\]
where $t_\epsilon = z/|\Lambda_\epsilon|$.
The eigenvalues (compare with \refeq{FP_eigs}) are simply
\[
e^{-\nu_k} = \frac{1}{|\Lambda_0|\Lambda_0^{k}} +
            \frac{1}{|\Lambda_1|\Lambda_1^{k}}
\,\, .
\]
Asymptotically the spectrum is dominated by the lesser
of the two fixed point slopes
$\Lambda=\Lambda_0$
(if $|\Lambda_0| < |\Lambda_1|$, otherwise $\Lambda=\Lambda_1$),
and the eigenvalues $e^{-\nu_k}$
fall off exponentially as $1/\Lambda^{k}$, just as in the
single fixed-\-point example above.

The proof that the \Fd\ for a general nonlinear 1-$d$  map
\refeq{1d_rep} is entire uses the expansion
\[ 
   \det(1-z\Lop) = \sum_{k \geq 0} (-z)^k \tr \left(\wedge^k \! \Lop\right)
\]
where $\wedge^k \Lop$ is the $k$'th exterior power of the operator $\Lop$.
For example,
\[ \wedge^2 \Lop (x_1 x_2,y_1 y_2) = \frac{1}{2!}
   \left[ \begin{array}{cc} \Lop(x_1,y_1) & \Lop(x_2,y_1) \\
               \Lop(x_1,y_2) & \Lop (x_2,y_2)
          \end{array} \right]
\]
so $\Tr (\wedge^2 \Lop) = \frac{1}{2!}
        \left( (\Tr \;\Lop)^2 - \Tr (\Lop^2)\right)$.
In a suitable polynomial basis $\phi_n(z)$ the operator
has an explicit matrix representation
\[ (\Lop \phi)_n(z) = \sum_{m = 0}^\infty \Lop_{mn}\phi_m(z)
\,.
\]
In the single fixed-point example \refeq{FP_eigs},
$\phi_n = y^n$, and $ \Lop $ is diagonal,
$\Lop_{nn}= \Lambda^{-n}/|\Lambda|$.

The proof proceeds by employing Cauchy complex contour integrals in order to
verify that the traces are indeed
given by the heuristic formula \refeq{tr_L}. Furthermore, from bounds on the
elements $\Lop_{mn}$ one calculates bounds
on  $\tr \left(\wedge^k \! \Lop\right)$
and verifies\rf{Ruelle76,frie,Rugh92}
that they again fall off as $\ExpaEig^{-k^2/2}$,
concluding that the $\Lop$ eigenvalues fall off exponentially
for a general Axiom A 1-$d$ map.
The simplest example of how the Cauchy formula
is employed is provided by a nonlinear inverse map
$\psi=f^{-1}$,
$s = \mbox{sgn}( \psi')$
\[ \Lop \phi(w)
  = \int d\!x \; \delta(w-f(x)) \phi(x)
  = s \; \psi'(w)\; \phi(\psi(w))
\,.\]
Assume that $\psi$ is a contraction of the unit disk, i.e.
\[ |\psi(w)| < \theta < 1 \ \ \ \mbox{and} \ \ \
   |\psi'(w)| < C < \infty
  \  \ \ \mbox{for} \ \ \ |w|<1
\,,
\]
and expand $\phi$ in a polynomial basis 
by means of the Cauchy formula
\[ \phi(z) = \sum_{n \geq 0} z^n \phi_n = \oint \Id \;\frac{\phi(w)}{w-z}
\,,\quad
\phi_n = \oint \Id\; \frac{\phi(w)}{w^{n+1}} \,.\]
In this basis, $\Lop$ is a represented by the matrix
\[ \Lop \phi(w) = \sum_{m,n} w^m L_{mn} \phi_n
\,,\quad
L_{mn} = \oint \Id \;\frac{s \; \psi'(w) (\psi(w))^n}{w^{m+1}} \]
Taking the trace and summing we get:
\[ \Tr \;\Lop = \sum_{n\geq 0} L_{nn} = \oint \Id\;
    \frac{s \; \psi'(w)}{w-\psi(w)} \]
This integral has but one simple pole at the unique fix point
$w^* = \psi(w^*) = f(w^*)$. Hence
\[ \Tr \;\Lop = \frac{s \; \psi'(w^*)}{1-\psi'(w^*)} =
    \frac{1}{|f'(w^*)-1|} \]
The requirement that map be analytic is needed to guarantee the
inequality
\[ |L_{mn}| \leq \sup_{|w|\leq 1} |\psi'(w)| \; |\psi(w)|^n
   \leq C \theta^n \]
which shows that finite $[N\times N]$
matrix truncations approximate the operator within an error
exponentially small in $N$.

We note in passing that for
1-$d$ repellers a diagonalization of an explicit
truncated $\Lop_{mn}$ matrix yields many more eigenvalues than
the cycle expansions\rf{CCR,Rugh92}. The reasons why one persists
anyway in using the periodic orbit theory are partially aestethic,
and partially pragmatic. Explicit $\Lop_{mn}$ demands explicit
choice of a basis and is thus non-invariant, in contrast to
cycle expansions which utilize only the invariant information
about the flow. In addition, we do not know how to construct
$\Lop_{mn}$ for a realistic flow, such as the 3-disk problem,
while the periodic orbit formulas are general and
straightforward to apply.

It is a relatively simple task to generalize the above arguments to
an expanding $d$-dimensional
dynamical system $f : M \rightarrow M$
with the Markov property
(for a more precise definition see ref.~\rf{Ruelle76}):
\\

\underline{Markov property:} One demands that $M$ can be divided into
$S$ subsets $\{I_0, I_1, \dots ,I_{S-1}\}$
such that either $f I_i \cap I_j = \emptyset$
or $I_j \subset f I_i$. The transition matrix takes
values $t_{ij} = 0$ or $1$, accordingly.

\underline{Expansion property:} Each inverse
 $\psi_{ij} : I_j \rightarrow I_i$
(defined when $t_{ij}=1$) is unique and a contraction.
\\

Depending on the smoothness of the functions $\psi_{ij}$ and the function
space considered, as well as how
one defines the contraction, one obtains stronger  or weaker results on
the spectrum of eigenvalues.
We shall restrict ourselves to the space of analytic functions,
and assume that there exists
a set of complex neighborhoods $D_i \supset I_i$
such that $\psi_{ij} : \mbox{Cl}(D_j) \rightarrow \mbox{Int}(D_i)$.
Mapping {\em closures} of domains into {\em interiors} is a useful
way of stating the contraction property.
The result is\rf{Ruelle76,frie,losal,Rugh92} that
the expansion coefficients $C_k$
fall off as \ $ 1/\ExpaEig^{k^{1 + 1/d}}$,
and the eigenvalues as \ $1/\ExpaEig^{k^{-1/d}}$.
Again, the results can be proved using a multinomial basis on each
domain and deducing from this an explicit matrix representation
for the operator.

\subsection{Hyperbolic maps}

{\bf Theorem [Rugh 1992]:} {\em The \Fd\ for hyperbolic analytic maps
is entire.}
\\

The proof, apart from the Markov property
which is the same as for the purely expanding case,
relies heavily on analyticity of the map
in the explicit construction of the function space and its basis.
The basic idea of the proof
is to view the hyperbolicity as a cross product
of a contracting map in the forward time and another contracting map in
the backward time.
In this case the Markov property introduced above has to be elaborated a bit.
Instead of dividing the phase space into intervals, one divides it
into rectangles. The rectangles should be viewed as direct product of
intervals (say horizontal and vertical), such that the forward map
is contracting in, for ex. the horizontal direction, while the inverse
map is contracting in the vertical direction.
For Axiom A systems the natural coordinate axes are given  by the
stable/unstable manifolds of the map. 
With the phase space divided into $S$ rectangles
$\{R_0, R_1, \dots ,R_{S-1}\}$,
$R_i = I^h_i \times I^v_i$
with complex extension $D^h_i \times D^v_i$, the hyperbolicity condition
(which at the same time guarantees the Markov property) 
can be formulated as follows:\\

\underline{Analytic hyperbolic property:}
Either $\,\, f R_i \cap \mbox{Int} (R_j) = \emptyset$, or
for each pair $w_h \in \mbox{Cl}(D^h_i)$, $z_v \in \mbox{Cl}(D^v_j)$
there exist unique analytic functions of $w_h,z_v$:
  $w_v = w_v (w_h,z_v) \in \mbox{Int}(D^v_i)$,
  $z_h = z_h (w_h,z_v) \in \mbox{Int}(D^h_j)$,
such that $f(w_h,w_v) = (z_h,z_v)$.
Furthermore, if $w_h \in I^h_i$ and $z_v \in I^v_j$, then
$w_v \in I^v_i$ and $z_h \in I^h_j$. (See fig.~\ref{hhr:rects}).
\\

What this means is that it is possible to replace coordinates $z_h,z_v$
at time $n$ by the contracting pair $z_h,w_v$, where $w_v$ is the contracting
coordinate at time $n+1$ for the inverse map.
Specifying the closure/interior of the sets
is a convenient way of defining hyperbolicity.

A map $f$ satisfying the above condition is called analytic hyperbolic
and the theorem 
states that the associated
Fredholm determinant is entire, and that the trace formula \refeq{tr_L}
(derived heuristically in ref.~\cite{AACI}) is correct.
We refer the reader to ref.~\cite{Rugh92} for the details of the proof.
The theorem applies also to hyperbolic analytic maps in $d$~dimensions and
smooth hyperbolic analytic flows in $(d+1)$~dimensions, provided that
the flow can be reduced to a map by suspension on a Poincar\'e section
complemented by an analytic ``ceiling" function\rf{bowen} which accounts
for a variation in the section return times. For example, if we take as
the ceiling function $g^t(x)=e^{sT(x)}$, where $T(x)$ is the time of the next
Poincar\'e section for a trajectory staring at $x$, we reproduce
the flow \Fd\ \refeq{cl_weig}.

Examples of analytic hyperbolic maps are provided by
small analytic perturbations of
the cat map (where the Markov partitioning is non-trivial\rf{deva}),
the 3-disk repeller, and the 2-d baker's map, the last
two examples to be discussed further below.

The proofs of discreteness of the classical spectra
have so far not been extended to
the semi-classical \qS s, with exception of the spaces of
constant negative curvature\rf{selberg}. The technical problem
is that the proofs require an evolution operator that is multiplicative
along the trajectory; composition of the semi-classical
Green's functions is not of that type, as every composition
of successive semi-classical paths requires
a further saddle-point approximation (see sect.~\ref{QUANTUM}).

\section{CLASSICAL AND QUANTUM DETERMINANTS}
\label{DETERMINANTS}

Though the only objective of this paper is to compare the
convergence of the semi-classical determinants \refeq{Fred_qm}
and \refeq{Selb_qm}, theoretical motivation
demands a plethora of related zeta functions and determinants.
In this section we introduce and define the additional determinants
and zeta functions that will be required in what follows.

\subsection{The \qS }
\label{QUANTUM}

The semi-classical periodic orbit theory for hyperbolic
flows was developed by
Gutzwiller in terms of traces of Van Vleck semi-classical Green's
functions\rf{gutbook}, and subsequently re-expressed in terms of determinants
by
Voros\rf{voros}.
In contrast to the sharp delta-function kernel for the classical
evolution, the {\em quantum evolution operator}
is the smeared-out Green's function kernel
\beq
G(q'',q';t)
\ee{quantGreen}
defined only on a half of the phase space (typically either
the spatial coordinates $q$, or the momentum coordinates $p$).
\Gt\ follows by replacing the
quantum Green's function by the semi-classical Van Vleck propagator
$K_c(q'',q';t)$, and evaluating the trace $\tr K_c(q'',q';t)$ by
saddle-point methods. The result is a periodic orbit formula
of topologically the same structure as the classical
trace \refeq{tr_L_cont}, but with different weights:
the semi-classical {\em \Gt}\rf{gutbook}:
\beq
\tr G(E)
             = \overline{g}(E) + \frac{1}{i\hbar}\sum_p T_p \sum_r^{\infty}
        {       e^{ -\frac{i}{\hbar} S_p(E) r + i \pi {m_p \over 2} r}
              \over |\det \left( {\bf 1}-{\bf J}_p^r \right)|^{1 \over 2} }
\,\, .
\ee{tr_Gutz}
Here $T_p$ is the $p$-cycle period,
$S_p$ its action, $m_p$ the Maslov index, and ${\bf J}_p$ is the
transverse Jacobian of the flow.
As in many applications the wave number $k$ is a more natural choice of
variable than the energy $E$, we shall henceforth replace
$E \rightarrow k$ in all semi-classical formulas.

For 2-$d$ flows the \Gt\ is of the form (see \refeq{Jac_exp})
\beq
\tr G(k) = \overline{g}(k) + \frac{1}{i\hbar}\sum_p T_p \sum_r^{\infty}
        {       e^{ -\frac{i}{\hbar} S_p(k)r + i \pi m_p r/2}
              \over |\Lambda_p^r|^{1/2} \left( 1- 1/\Lambda_p^r \right) }
\,\, .
\ee{tr_Gutz_2d}
and the corresponding determinant \refeq{der_det} is the
2-$d$ {\em \qS}
\bea
\Zqm(k) &=& \exp\left( - \sum_p \sum_r^{\infty} {1 \over r}
        {       t_p^r
              \over \left( 1- 1/\Lambda_p^r \right) }
                   \right)
\,,
        \continue
t_p     &=& z^{n_p} { e^{-\frac{i}{\hbar} S_p + i \pi m_p/2}
         \over {\sqrt{|\Lambda_p|} }}
\,\, .
\label{t_p_quant}
\eea
$z$ is a book-keeping variable that
keeps track of the topological cycle length $n_p$, used to
expand zeta functions and determinants
(see sect.~\ref{CYCLE}).
Unlike the \cFd, which is exact, the \qS\
is the leading term of a
semi-classical approximation, and the size of corrections
to it remain unknown\rf{wirzba}.

The \Gt, apart from the quantum and Maslov phases, differs from
classical trace formula in two aspects. One is the volume
term $\overline{g}(E)$ in \refeq{tr_Gutz} which is missing from our
version of the classical trace formula. While an overall pre-factor
does not affect the location of zeros of the determinants, it might
play a role in relations such as functional equations for zeta functions.
The other difference is that the quantum kernel leads to a
square root of the cycle Jacobian determinant, a reflection of
the relation probability~=~amplitude$^2$.
The $1/\sqrt{\det(1- {\bf J}_p)}$ weight leads in turn
to the product representation \refeq{Selb_qm}
\beq
\Zqm(k) = \prod_p \prod_{k=0}^\infty
   \left( 1 - t_p / \Lambda_p^k  \right)
\,\, ,
\ee{2d_quant}
which differs from the \cFd\ \refeq{cl_weig} by missing exponent $k+1$.

\subsection{The \qFd}

The conjecture tested in this paper asserts that
one may replace the \qS\ \refeq{2d_quant} by the
{\em \qFd} \refeq{Fred_qm}, {\em ie.} the
\Fd\  \refeq{cl_weig} with the {\em quantum} weights $t_p$,
without disturbing the leading semi-classical eigenvalues,
but improving the convergence of cycle expansions used in
evaluating the spectrum, and revealing part of spectra inaccessible to
the \qS\ (see sect.~\ref{3-DISK}  for an example).

The form of the quantum weight \refeq{t_p_quant} suggests that
the quantum evolution operator should be approximated by a
classical evolution operator with a quantum weight:
\[
\Lop^t (y,x) =
                  \delta ( y - f^t(x))
        \sqrt{|\Lambda^t(x)|} e^{-\frac{i}{\hbar} S^t(x) + i \pi m_p(x)/2}
\,\, .
\]
As explained in sect.~\ref{ENTIRE},
this operator is not multiplicative along the trajectory,
and consequently does not satisfy the assumptions required by
the theorems that guarantee that a \Fd\ is entire. Nevertheless,
our numerical results support the conjecture
that the $|\Lambda|^{1/2}$ weighted determinant
has a larger domain of analyticity than the commonly used \qS\
and that some related determinant might even be entire.

\subsection{The \dzeta}

The Ruelle or the \dzeta\cite{ruelle} is defined by
\beq
1/\zeta = \exp\left( - \sum_p \sum_{r=1}^\infty {1 \over r} t_p^r
                       \right)
\,\, ,
\ee{dynzeta}
and has the Euler product representation
\beq
1/\zeta =  \prod_p { ( 1- t_p ) }
\,\,,
\ee{zet}
where the product is over all prime cycles $p$,
and the cycle weight $t_p$ depends on the average computed;
the simplest example is the weight \refeq{cl_weig} used in
computation of escape rates and correlation spectra.
(\ref{zet}) also yields the leading semi--classical {\em quantum}
resonances, if $t_p$ is the quantum weight \refeq{t_p_quant}
associated with the cycle $p$.

Historical antecedents of the \dzeta\ are the fixed-point counting
functions introduced by Artin-Mazur and Smale\rf{smale}, and the
determinants of transfer operators of statistical mechanics\rf{sinai}.
While $1/\zeta$ is the natural object in these applications, in
dynamical systems theory \dzeta s arise naturally only for
piecewise linear mappings; for smooth flows the natural object for
study of classical and quantal spectra are the determinants
introduced above. However, \dzeta s will here be useful objects
in relating various determinants and explaining the geometrical
meaning of curvature expansions.

\subsection{Weighted \Fd s}

The theorems of sect.~\ref{ENTIRE} apply not only to
the evolution operator \refeq{TransOp}, but also to more general
evolution operators multiplicative along the flow. In particular,
transport of $k$-forms transverse to the flow is given by the
exterior products of the Jacobian  \refeq{jacob} of the flow
\beq
\Lop^t_{(k)} (y,x) =  \delta ( y - f^t(x)) \cdot  \wedge^k {\bf J}
\,\, .
\ee{TransWedge}
In $d$ dimensions (transverse to the flow), the non-vanishing
exterior powers are ${\bf J}$, $\wedge^1 {\bf J}$, $\wedge^2 {\bf J}$,
$\dots$, $\wedge^d {\bf J}$.
The $k=0$  case describes scalar density transport \refeq{TransOp};
$k=1$  describes vector transport (utilized in evaluation of
strange set stabilities\rf{AACI,AACII,CCR} and
fast dynamo rates\rf{dynamo});
$k=2$ describes $dx \wedge dy$ area transport, and so on.
$\Lop^t_{(k)}$ is a [$d^k$$\times$$d^k$] matrix; for example,
in the explicit index notation $\wedge^2 {\bf J}$ is given by the
antisymmetric exterior product
\[
(\wedge^2 {\bf J})_{ac,bd} = {1 \over 2}
        \left(\delta^e_a \delta^f_c - \delta^f_a \delta^e_c\right)
        J_{eb} J_{fd}
\,.
\]
As $\left( \wedge^k {\bf A} \right) \left( \wedge^k {\bf B} \right)
        = \wedge^k {\bf AB} $,
$\Lop^t_{(k)}$ are multiplicative operators along the flow, so
if $\det(1-\Lop)$ is entire, $\det(1-\Lop_{(k)})$ are also entire.
The trace formula for $\Lop_{(k)}$ is
\beq
\tr \Lop_{(k)}(s)
             = \sum_p T_p \sum_r^{\infty}
        { \tr \left( \wedge^k {\bf J}_p^r \right) e^{sT_pr}
           \over |\det \left( {\bf 1}-{\bf J}_p^r \right)| }
\,\, ,
\ee{tr_L_k}
where $\tr {\bf J} = \Lambda_1 + \Lambda_2 + \dots + \Lambda_d$,
$\tr \wedge^2 {\bf J} = \Lambda_1 \Lambda_2 +
                \Lambda_1 \Lambda_3 +\dots + \Lambda_{d-1}\Lambda_d$,
and so on.
The corresponding \Fd s are:\\
{\em 1-$d$ case}:
\[
\det(1-\Lop_{(1)}) =  \prod_p  \prod_{k=0}^\infty
                        \left( 1 -   t_p /\Lambda_p^{k-1} \right)
\, ,
\]
{\em 2-$d$ Hamiltonian case}:
\bea
\det(1-\Lop_{(1)}) &=&  \prod_p  \prod_{k=0}^\infty
                        \left( 1 -   {t_p \over \Lambda_p^{k-1}} \right)^{k+1}
                        \left( 1 -   {t_p \over \Lambda_p^{k+1}} \right)^{k+1}
                \continue
\det(1-\Lop_{(2)}) &=&  \prod_p  \prod_{k=0}^\infty
\left( 1 -   t_p /\Lambda_p^{k} \right)^{k+1} = \det(1-\Lop)
\, ,
\nnu
\eea
where we have used the Hamiltonian volume conservation in
$\tr \wedge^2 {\bf J}_p = \Lambda_p \cdot 1/\Lambda_p =1$.

In order to discuss relations between various determinants and
zeta functions, it is convenient to also define following
weighted \dzeta s:
\beq
1/\zeta_k=\prod_p (1-t_p/\Lambda_p^{k})
         = \exp\left( - \sum_p \sum_{r=1}^\infty {1 \over r}
                      {t_p^r \over  {\Lambda_p^{kr}} } \right)
\,\, ,
\ee{wdynzeta}
and weighted \Fd s
\beq
F_k = \exp\left( - \sum_p \sum_{r=1}^\infty {1 \over r}
                 { (t_p / \Lambda_p^k)^r
                  \over
                   (1-1/\Lambda^r_p)^2
                 }
        \right)
\,\, .
\label{F_k}
\eeq
($F_0 = F$).
For 2-dimensional Hamiltonian flows the \cFd\
\refeq{cl_weig} can be written as
\beq
F
=\prod_{k=0}^{\infty} 1/\zeta_k^{k+1}
\,\, .
\ee{2d_Fred}
$F_k$ can be interpreted as the \Fd\   $\det(1-\Lop_k)$
of the weighted evolution operator
\[
\Lop_k^t (y,x) =   {g^t(x) \over \Lambda^t(x)^{k} }
                  \delta ( y - f^t(x))
\,\, ,
\]
where $\Lambda^t(x)$ is the expanding eigenvalue of the
Jacobian transverse to the flow,
and $g^t(x)$ is any analytic weight function multiplicative along the
trajectory. As discussed in sect.~\ref{ENTIRE}, there is no
guarantee that $\det(1-\Lop_k)$, the \Fd\ for this operator,
should be entire for $k \neq 0$.

\subsection{Relations between determinants and zeta functions}
\label{RELATIONS}

Ruelle\rf{Ruelle76} has observed that the elementary identity
for $d$-dimensional matrices
\[
1= {1 \over \det(1-{\bf J})}
        \sum_{k =0}^d (-1)^k \tr \left(\wedge^k {\bf J}\right)
\,,
\]
inserted into the exponential representation \refeq{dynzeta} of the
\dzeta, relates the \dzeta\ to weighted \Fd s.
In one dimension this identity
\[
1= {1 \over {(1-1/\Lambda)}}
   - {1 \over {\Lambda}} {1 \over {(1-1/\Lambda)}}
\]
expresses the \dzeta\ as a ratio of two \Fd s
\beq
1/\zeta = {\det(1-\Lop) \over \det(1-\Lop_{(1)}) }
\ee{Fred_ratio}
and shows that $1/\zeta$ is meromorphic, with poles given by
the zeros of $\det(1-\Lop_{(1)})$.
In refs.~\cite{CPR90,AACII}
heuristic arguments were developed
for 1-dimen\-si\-on\-al mappings to explain how the poles of individual
$1/\zeta_k$ cancel against the zeros of $1/\zeta_{k+1}$, and thus
conspire to make the corresponding \Fd s entire.
Numerical checks verify both the heuristic arguments,
and the formula \refeq{Fred_ratio}.

For 2-dimensional Hamiltonian
flows this yields
\[
1/\zeta = {\det(1-\Lop) \det(1-\Lop_{(2)}) \over \det(1-\Lop_{(1)}) }
        = {F^2 \over F_{1} F_{-1} }
\]
This establishes that $1/\zeta$ is meromorphic in 2-$d$ as well,
but the relation is not particularly useful for our purposes.
Instead we insert the identity
\[
1= {1 \over {(1-1/\Lambda)^2}}
   -{2\over \Lambda} {1 \over {(1-1/\Lambda)^2}}
   + {1 \over {\Lambda^2}} {1 \over {(1-1/\Lambda)^2}}
\]
into the exponential representation (\ref{wdynzeta}) of $1/\zeta_k$,
and obtain
\beq
1/\zeta_k = { {F_k F_{k+2}} \over {F_{k+1}^2}}
\,\, .
\ee{doub_pole}
Even though we have no guarantee that $F_k$ are entire, we
do know (by arguments explained in the next section) that the
upper bound on the leading zeros of $F_{k+1}$ lies strictly
below the leading zeros of $F_{k}$, and therefore we expect
that for 2-dimensional Hamiltonian flows the
\dzeta\  $1/\zeta_k$ has generically a {\em double} leading pole
coinciding with the leading zero of the $F_{k+1}$
\Fd.  This might fail if the poles and leading eigenvalues come in
wrong order, but we have not encountered such situation in our numerical
investigations.
This result can also be stated as follows: the theorem that establishes that
the \cFd\  (\ref{2d_Fred})
is entire, implies that the
poles in $1/\zeta_k$ must have right multiplicities in order
that they be cancelled in the $ F = \prod 1/\zeta_k^{k+1}$ product.

Both the \qFd\  and the
quantum zeta function yield  the same leading zeros, given by $1/\zeta_0$.
They differ in non-leading zeros (zeros with larger
imaginary part of the complex energy), but as
the \qS\ (\ref{2d_quant}) is only the leading term of a
semi-classical approximation, with the size of corrections unknown,
the physical significance of these non-leading zeros remains unclear.

\subsection{Abscissa of absolute convergence}
\label{ABSCISSA}

Consider the ``thermodynamic" approximation \refeq{thermo}
in the case of the \Gt\ \refeq{tr_Gutz_2d}:
\[
\tr G(k) \, \approx \,
\sum_p T_p \sum_r^{\infty}
        {       e^{ -\frac{i}{\hbar} S_p(k)r + i \pi m_p r/2}
              \over |\Lambda_p^r|^{1/2}}
\,\,.
\]
This approximation omits the non-leading $\approx 1/\Lambda_p$
terms that vanish in the $t \rightarrow \infty$ limit and do not
affect the leading eigenvalues.
If the phases conspire to partially cancel contributing terms,
the sum diverges for a larger value of $\mbox{Im}(k)$.
The abscissa of absolute convergence in the complex $k$ plane
is obtained by maximizing the sum, {\em ie}. replacing all terms
by their absolute values:
\[
\tr G(k) \, \approx \,
\sum_p T_p \sum_r^{\infty}
        {       e^{ T_p\mbox{\footnotesize Im}(k) r}
              \over |\Lambda_p^r|^{1/2} }
\,\,.
\]
(we have for simplicity taken $S_p$ to be the action for
billiards, $S_p/\hbar = T_p k$).
Value of $\mbox{Im}(k)$ for which this sum diverges determines
the abscissa of absolute convergence.
In the case of the \Gt\ \refeq{tr_Gutz_2d}
this leads to the disappointing result
that all eigenvalues of interest are outside the domain of
convergence.  However, for determinants and zeta functions
the eigenvalues are given by locations of the zeros,
and the analyticity domain is larger.
We can also use determinants to accurately estimate this
abscissa of absolute convergence, by replacing all cycle weights
$t_p$ in \refeq{cl_weig} by their absolute values.

To evaluate the abscissa of absolute convergence of the
\qS\ we first note that
inserting the identity
$ \,
1 = (1 - 1/\Lambda_p^{r}) / (1 - 1/\Lambda_p^{r})
\, $
into the exponent of \qS\
\refeq{t_p_quant},
one obtains the following relation between the \qS\ and
the \qFd:
\[
      \Zqm (k) = { \Fqm(k) \over F_{1 \over 2}(k)}
\,,
\]
where
\beq
F_{1 \over 2}(k) = \exp \left( -\sum_p \sum_{r=1}^{\infty} {1 \over r}
        {  t_p^r \over \Lambda_p^r (1 - 1/\Lambda_p^{r})^2 }
                                \right)
\,.
\ee{F1_2}
The radius of convergence of $\Zqm(k)$ is therefore determined
by the leading zeros of $F_{1 \over 2}(k)$. To
estimate the upper bound on Im$(k)$ for the zeros of $F_{1 \over 2}$,
we omit all signs and phases in the weights in $F_{1 \over 2}$:
\beq
\hat{F}_{1 \over 2}(k) =  \exp\left( -\sum_p \sum_{r=1}^{\infty} {1 \over r}
        {  |t_p^r| \over |\Lambda_p^r| (1 - 1/\Lambda_p^{r})^2 }
                                \right)
\,,
\ee{F1_2_abs}
and compute its leading zero at Re$(k) =0$. An example is given in
sect.~\ref{3-DISK}.

\subsection{The \qS\ puzzle: a classical determinant?}

On the basis of close analogy between the classical and the quantum zeta
functions, it has been hoped\rf{CHAOS92} that for
Axiom A systems the \qS s (\ref{2d_quant})
should also be entire.
This hope was dashed by Eckhardt and Russberg\rf{ER92} who have
established by numerical studies that the
\qS s for 3-disk repeller have poles.
They find numerically that
in the \qS\ (\ref{2d_quant}) product representation
$1/\zeta_0$ has a
double pole coinciding with the leading zero of $1/\zeta_1$.
Consequently $1/\zeta_0$, $1/\zeta_0 \zeta_1$ and $\Zqm$ all have the
same leading pole, and coefficients in their cycle expansions fall off
exponentially with the same slope. Our numerical tests on the 3-disk
system (see sect.~\ref{3-DISK}) support this conclusion.

Why should $1/\zeta_0$ have a {\em double} leading pole? The
double pole is not as surprising as it might seem at the first glance;
indeed, for the \cFd\ \refeq{cl_weig}
$1/\zeta_0$ must have a double pole in order to cancel the leading
double zero of the $(1/\zeta_1)^2$ factor. In other words,
the numerics indicates that the semi-classical determinants and zeta
functions have analyticity properties characteristic of {\em classical}
determinants.
As {\em a priori} any determinant or zeta function which yields the
same {\em leading} semi-classical resonances is equally good, one
is lead to pose the conjecture tested in this paper:
\qFd s are preferable to \qS s, as they yield the same leading
spectrum, but with better analyticity properties.

As we have no proof, we resort to numerical tests of the conjecture;
but first we need to briefly explain how such determinants are
investigated numerically.

\section{CYCLE EXPANSIONS}
\label{CYCLE}

A {\em cycle expansion}\cite{cycprl}
is a series representation of
a zeta function or a \Fd,
expanded as a sum over {\em pseudo-cycles}, products of
prime cycle weights $t_p$,
ordered by increasing cycle length and instability.
The products (\ref{2d_quant}), (\ref{zet}) are
 really only a shorthand notation
for zeta functions and determinants - for example, the zeros of the individual
factors in infinite products (\ref{2d_quant}), (\ref{zet}) are
{\em not} the zeros of the corresponding zeta functions
and determinants, and
convergence of such objects is far from obvious.

\subsection{Curvature expansions}

{\em Curvature expansions} are based on the
observation\cite{cycprl,AACI} that the motion in
dynamical systems with finite grammar is
organized around a few {\em fundamental} cycles; more precisely,
that the cycle expansion of the \dzeta\ (\ref{zet})
allows a regrouping of
terms into dominant {\em fundamental} contributions $t_f$ and
decreasing {\em curvature} corrections $c_n$:
\beq
   1/\zeta = 1 - \sum_f t_f - \sum_n c_n
\,.
\label{cycexp}
\eeq
The fundamental cycles $t_f$ have no shorter approximants;
they are the ``building blocks'' of the dynamics
in the sense that all longer orbits can be approximately
pieced together from them.  In piecewise linear approximations
to the flow, $1/\zeta$ is given by the determinant for a finite
Markov transition matrix, and all $c_n$ vanish identically.
Hence the designation ``curvatures"; size of $c_n$ is an indication
how far the flow is from a piecewise linearization.

A typical curvature term in
(\ref{cycexp}) is a {difference} of a long cycle $\{ab\}$
and its shadowing approximation by shorter cycles
$\{a\}$ and $\{b\}$:
\[
t_{ab}-t_a t_b = t_{ab} (1 - t_a t_b / t_{ab})
\]
The orbits that follow the same symbolic dynamics, such as $\{ab\}$
and the ``pseudo orbit" $\{a\} \{b\}$, lie close to each other, have
similar weights, and for longer and longer orbits the
differences are expected to fall off rapidly.
For systems that satisfy Axiom A requirements,
such as the 3-disk repeller, curvature expansions converge
very well\cite{eck}.
It is crucial that the curvature expansion is grouped (and
truncated) by topologically related cycles and pseudo-cycles;
truncations that ignore topology, such as inclusion of all cycles with
$T_p < T_{max}$, will contain un-shadowed orbits, and exhibit a
mediocre convergence compared with the curvature expansions.

\subsection{\Fd\ cycle expansions}

While for the \dzeta\ cycle expansions the shadowing is easy to
explain, the resulting convergence is not the best achievable;
as explained above, \Fd s are expected to be entire, and
their cycle expansions should converge faster than exponentially.
The \Fd\ cycle expansions are somewhat more complicated
than those for the \dzeta s. We expand the exponential representation
\refeq{Z(s)} of $F(s)$ as a multinomial in prime cycle weights $t_{p}$
\bea
F_p &=& 1 - \sum_{r=1}^\infty
       \frac{1}{r}
       \frac{t^r_p}{|\det \left[ {\bf 1}-{\bf J}_p^r \right]|}
        + \frac{1}{2} \left(\dots \right)^2
- \dots .
                \continue
    &=& \sum_{r=1}^\infty C_{p^k} t_p^k
\, .
\nnu
\eea
This yields the cycle expansion for $F(s)$:
\bea
F(s) &=&
\sum_{k_1 k_2 k_3 \cdots = 0}^{\infty}
\tau_{p_1^{k_1} p_2^{k_2}  p_3^{k_3} \cdots }
        \continue
\tau_{p_1^{k_1} p_2^{k_2}  p_3^{k_3} \cdots } &=&
\prod_{i=1}^\infty C_{{p_i}^{k_i}} t_{p_i}^{k_i}
\,\, ,
\nnu
\eea
where the sum goes over all pseudo-cycles.
The coefficients have a simple form only in 1-$d$, given by the
Euler formula \refeq{Euler}. Expansions for the 2-$d$ case are
discussed in refs.~\cite{losal,ER92}.

In practice we do not do anything as complicated.
We evaluate numerically the exponent in \refeq{Z(s)} as a power series in the
book-keeping variable $z$, truncated to maximal cycle length $N$.
The we expand the exponential, keeping terms up to $N$ and obtain
\beq
F_N(s,z) = \sum_{k=0}^{N} C_{k}(s) z^k
\,\, ,
\ee{cyc_exp}
and similarly for the \qS\ and the \dzeta s.
In the final evaluation $z$ is set to $z=1$,
but the organization by powers of $z^k$ is crucial to
the convergence of cycle expansions. For example, for
1-$d$ Axiom A mappings
$C_k(s) \approx C^{-k^2}$  for any $s$ - this super-exponential convergence
is precisely the reason why the variable $z$ was introduced
in the first place.
The zeros of $F_N(s,1)$ are determined by standard methods, such as
the Newton-\-Raphson
algorithm.

Russberg and Eckhardt\rf{ER92} have generalized
curvature expansions to \Fd s; they work to a point, but are less
powerful than the general theorems on Axiom A flows.
For example, they predict that the \cFd\ could have poles,
in violation of the theorem explained above.

\subsection{Convergence of cycle expansions}

It is fairly easy to establish that for Axiom A systems
the trace formulas converge exponentially with the number of
cycles included. As explained in sect.~\ref{ABSCISSA},
the trace formulas are not absolutely convergent where you need them,
and in addition, shadowing of longer orbits by nearby
pseudo-orbits is not implemented, so we will not use
trace formulas at all. However, it should be noted that
for systems other than Axiom A, we do not know how to
improve convergence by shadowing cancellations, or define
determinants that are guaranteed to be entire, and it is
still possible that for generic systems
determinants do not converge any better than traces.

For \dzeta s geometrical estimates\rf{AACI} imply that for
Axiom A systems the curvature expansion coefficients fall off exponentially,
$C_k \approx {\tilde{C}}^k$, and the expansion sums up to a pole
\[
      \sum_{n=0}^{\infty} C_k z^k  \simeq  \sum_{n=0}^{\infty}
({\tilde C}z)^k  =  \frac{1}{1 - {\tilde C}z}.
\]
Such poles are expected from Ruelle's relation \refeq{Fred_ratio} between
\dzeta s and \Fd s, and are indeed observed numerically\rf{AACII}.
Convergence of \dzeta s cycle expansions can be accelerated by a
variety of numerical methods, but both on theoretical grounds and
in practice, the preferred alternative is to use \Fd s instead.

As shown in sect.~\ref{ENTIRE}, for Axiom A maps
the coefficients in the \cFd\ $F(z)=\sum_n C_n z^n$ expansions
fall off faster than exponentially, as
$C_n \approx \Lambda^{-n^{2}}$ for 1-$d$ maps, as
$C_n \approx \Lambda^{-n^{3/2}}$ for 2-dimensional maps / 3-dimensional
flows, and in $d$ dimensions as
$\Lambda^{-n^{1+1/d}}$.
These estimates are confirmed by the
numerical tests of ref.~\cite{Rugh92},
numerical results for the 3-disk repeller ref.~\cite{ER92},
as well as the numerical results presented below.

\section{NUMERICAL RESULTS}
\label{NUMERICAL}

In this section we present the evidence that
the quantum \Fd\ is numerically as convergent as the
\cFd, in contrast to the \qS\ which has a finite radius of
convergence.

\subsection{3-disk resonances}
\label{3-DISK}

Following refs.~\cite{eck,ER92,gasp,GA}, we start by performing our
numerical tests on the 3-disk repeller.
The 3-disk repeller is the simplest physical realization of
an Axiom A system, particularly convenient for numerical investigations.
The methods that  we use to extract periodic orbits,
their periods and their stability eigenvalues are
described in ref.~\cite{CERRS}.
For billiards the cycle weight $t_p$ required for
evaluation of the classical escape rates and correlation spectra
is given by \refeq{cl_weig}.
The action $S_p$ is proportional to the cycle period $T_p$,
and the Maslov index changes by $+ 2$ for each disk bounce, $m_p = 2 n_p$,
so the quantum weight~(\ref{t_p_quant}) is given by
\beq
      t_p = (-1)^{n_p}\frac{e^{-ikT_p}}{\sqrt{|\Lambda_p|}} z^{n_p}
\, ,
\ee{q_act_bill}
where $k=(\mbox{momentum})/2\pi$ is the wave-number, and we take
velocity~=~1, mass~=~1.

Cycle expansion \refeq{cyc_exp} coefficients $|C_n|$
for different determinants and zeta functions are plotted
in figs.~\ref{fig1}~and~\ref{fig2}
as function of the topological cycle length $n$.
Zeta functions exhibit exponential falloff,
implying a pole in the complex plane,
while both the classical and the \qFd s  appear to  exhibit
a faster than exponential falloff, with no indication of a finite
radius of convergence within the numerical validity of our cycle
expansion truncations.

In particular, the \qFd\  enables us
to uncover a larger part of the quantum resonances than what
was hitherto accessible by means of the  \dzeta s\rf{GA,ER92,CERRS}.
The eye is conveniently guided to the zeros
by means of complex $s$ plane contour plots,
with different intervals of the absolute value of the
function under investigation assigned different colors;
zeros emerge as centers of elliptic neighborhoods of
rapidly changing colors.
Detailed scans of the whole area of the complex
$s$ plane under investigation and
searches for the zeros of classical and \qFd s,
fig.~\ref{qf.cl.zeros},
reveal complicated patterns of resonances, with the classical
and the semi-classical resonance patterns surprisingly similar.
It is known\rf{eck} that the leading semi-classical resonances are very
accurate approximations to the exact quantum resonances; the
semi-classical resonances further down in the $s$ complex plane in
fig.~\ref{qf.cl.zeros} have not yet been compared with the exact
quantum values. It would be of interest to check whether also all
\qFd\ resonances correspond to the
true quantum resonances.

An interpretation of the resonance spectrum of the
\cFd\ is given in ref.~\cite{GA}, where the resonances are
related to the oscillations in $N(t)$, the number
of particles that have not escaped by the time $t$,
with the basic frequency
$
      \overline{\omega}  = 2\pi /\overline{T}
$
given by the inverse of $\overline{T}$, the mean flight time between the disks.
$\overline{\omega}$
yields the mean spacing of the resonances along the imaginary $k$ axis.
In the 3-disk system there are two fundamental frequencies,  $\omega_0$ and
$\omega_1$, determined by the inverse periods of the
two fundamental cycles $\overline{0}$ and $\overline{1}$.
Corresponding beat frequency\rf{EPRI} $f = 2\pi /(T_1 - T_0)$ is clearly
visible in fig.~\ref{qf.cl.zeros}.
A rough measurement of the period
of the beats in fig.~\ref{qf.cl.zeros} yields some 23.7 units along the
real $k$ axis, to be compared to $2\pi /(T_1 - T_0)= 23.4$.

Effect of sub-dominant resonances on the measurable spectra are
exponentially small, and presumably of little physical interest. We
investigate them in detail here mostly in order to demonstrate that
our determinants indeed exhibit better convergence than the \qS.

Contour plots are also helpful in comparing the domain of
convergence of the \Fd\ to that of the \qS.
As can be seen from fig.~\ref{qfselbrast}, the \qFd\ can
be continued considerably farther down in the complex $k$ plane,
in contrast to the \dzeta\ scans such as those given in ref.~\cite{GA}.
While the zeta functions clearly exhibit a finite radius of convergence,
in agreement with the arguments of sect.~\ref{DETERMINANTS}, both the
\cFd\ and the \qFd\ behave as entire functions.
We compute the abscissa of absolute convergence for the \qS\ by
means of \refeq{F1_2_abs}; for the case at hand we obtain
the leading zero at $k_{ac}= 0.0 - i \, 0.699110157151 \dots$.
Indeed, comparison of the contour
plots of fig.~\ref{qfselbrast} shows that no feature of the
\qS\ countor plot below $k_{c}$ is significant. Interestingly
enough, the apparent border of \qS\ convergence in
fig.~\ref{qfselbrast} seems to
coincide with $\mbox{Re}(s) = 0$, $\mbox{Im}(s) = -1.09653395\dots$,
the zero obtained from $F_{1/2}(k)$
by removing quantum phases,
$t_p \rightarrow |t_p|$, but keeping the eigenvalue $\Lambda_p$
sign. in eq.~\refeq{F1_2}.

As discussed above, for 1-$d$ systems
the pole of $\zfct{0}$ coincides with the
leading zero of $\zfct{1}$, and the resulting product remains finite and has a
zero close\cite{AACI,CCR} to the leading $\zfct{1}$ zero.
In simple examples, such as the symmetric 1-dimensional tent map repeller,
the non-leading~/~leading zeros of the classical
$F_0$~/~$F_1$ maps are identical.
This suggests that some of the {\em non-leading} zeros of $F$ are
shadows of the $\zfct{1}$ zeros and hence
likely to lie close to the {\em leading} zeros of $F_1$.
For example, the non-leading resonance
$k=0.9915231008+i\,12.5163342128$ of the \cFd\ (see fig.~\ref{qf.cl.zeros}),
while distinct from the leading resonance of $F_1$ at
$k=0.99197582677+i\,12.5029206443798$, belongs to a family
of resonances that all lie very close to the leading $F_1$
resonances. We are confident that such resonances are distinct,
as their cycle expansions converge super-exponentially
to all digits listed above,
in agreement with the general theory.

The $F$ spectrum that is not echoed by the $F_1$ spectrum
can be isolated by ``subtracting" the $F_1$ spectrum,
{\em ie}. removing those resonances from the
$F$ spectrum that lie closer than some $\epsilon$
to a $F_1$ resonance. The resulting spectrum is
depicted in fig.~\ref{F_0-F_1}.

What is particularly interesting about this
spectrum is that with the resonances associated with the $F_1$
spectrum removed, families that connect non-leading resonances
at Re$(k)=0$ with the leading part of the resonance spectrum for
larger Re$(k)$ are clearly visible in fig.~\ref{F_0-F_1}.
This makes it rather clear that
the evaluation of leading resonances
for large Re$(k)$ requires inclusion of longer cycles,
the same that are required to control the spectrum for large negative
Im$(k)$ at Re$(k)=0$. This is the conundrum of semi-classical cycle
expansions; while semi-classical intuition implies that the
\qS\ should be applicable for large Re$(k)$, in practice the
semi-classical cycle expansions work best
for the bottom of the spectrum.

\subsection{H\'enon mapping as a normal form for a flow}
\label{MODELLING}

Our next example is a model of a 3-dimensional flow
$\dot{\bf x}={\bf F}({\bf x}),\;\; {\bf x}=(x_1,x_2,x_3)$.
We assume that the flow of interest is recurrent,
and that given a convenient Poincar\'e section coordinatized by
coordinate pair $(x,y)$, the flow can be described
by a 2-dimensional Poincar\'e map
\beq
P: \left\{ {
         x' = f_1(x,y)
         \atop
         y' = f_2(x,y)
         } \right.
\,,
\ee{Section}
together with the ``ceiling"\rf{bowen} function $T(x,y)$ which gives the
time of flight to the next section for a trajectory starting at $(x,y)$.
A trajectory $(x_1,y_1),(x_2,y_2),(x_3,y_3),\dots$ induces
a sequence of flight times
$T_1,T_2,T_3,\dots$, $T_k=T(x_k,y_k)$. This sequence can be used to
construct an embedding space, with the pair $(T_k,T_{k+1})$
serving as the embedding coordinates.
The embedding theories\rf{tak} imply that the
sequence $(x_k,y_k)$ generated by the original
map is dynamically equivalent with the sequence $(T_k,T_{k+1})$
(provided that the coordinate transformation is non-singular,
for example $T =$const is excluded),
and that the mapping
\begin{equation}
M:\;\;(T_k,T_{k+1})\rightarrow \;(T_{k+1},T_{k+2})\label{M}
\end{equation}
gives us the full invariant information about the dynamical system.
In particular, both the return times $T_k$ and
the stability eigenvalues of periodic orbits computed from
(\ref{M}) are equal to those of the original map \refeq{Section}.
While the  functional form of (\ref{M}) can be complicated in actual
implementation, we shall use here these ideas only as a
motivation for the following rather simple model:

Assume that the Poincar\'e map of the flow can be modelled
by a single analytic map for the whole $T$ range of interest:
\begin{equation}
T_{k+1}=G(T_{k},T_{k-1}).
\end{equation}
In general $G$ can be a complicated function, but the
essential properties of a continuous flow can be
modelled by the H\'enon map, which we take as a local
normal form (up to quadratic terms) of the time mapping.
In the Hamiltonian case, the form of the H\'enon map is
\bea
T_k &=& {\overline{T}} +\alpha x_k\label{times}
                \continue
x_{k+1}&=&1-ax^2_k-x_{k-1}
\,.
\label{time_spread}
\eea
${\overline{T}}$ is essentially the mean time of flight between
Poincar\'e sections which we shall set to ${\overline{T}}=1$,
and the parameter $\alpha$ allows us to
choose a narrow or a broad return time distribution
around the mean return time.
For $\alpha=0$ the flow reduces to the usual H\'enon map,
with constant time between the iterations. For example, the
$ R : a = 6 : 1 $ 3-disk repeller has a rather thin
repeller, and can be roughly fit with $a \approx 20$ in
\refeq{time_spread}.

We relate this hypothetic flow
to a semi-classical system by interpreting
the flight times ($\ref{times}$) as the lengths of
segments of billiard trajectories, with the action
of the periodic orbit $p$ given by
\beq
{1 \over \hbar} S_p(k) =k \sum_{i=1}^{n_p} T_i =
        k (n_p  + \alpha \sum_{i=1}^{n_p} x_{p,i})
\,,
\label{acti}
\eeq
where $x_{p,i}$ is the coordinate of the $i$-th cycle point of
prime cycle $p$, and $k$ is the wave number.
For billiards, the Maslov index $m_p$ is taken equal to 0 for cycles with
positive  stability eigenvalue $\Lambda_p$, and 1 for $\Lambda_p$
negative. We take unit velocity, so $T_p=L_p$.

For the complete repeller case (all binary sequences are realized),
the cycles are evaluated as follows.
According to
(\ref{times}),  the coordinates of a periodic orbit of
length $n_p$ satisfy the  equation
\beq
x_{p,i+1}+x_{p,i-1}=1 -  a x_{p,i}^2 \, ,
\quad i=1,...,n_p
\,,
\label{pere}
\eeq
with the periodic boundary condition
$x_{p,0}=x_{p,n_p}$.
In the complete repeller case, the H\'enon map is a realization
of the Smale horseshoe, and the symbolic dynamics has a very simple
description in terms of the
binary alphabet  $\epsilon = 0,1$,
$\epsilon_{p,i}=(1+S_{p,i})/2$, where $S_{p,i}$ are the
signs of the corresponding cycle point coordinates, $S_{p,i}=\sign{x_{p,i}}$.
We start with a preassigned sign sequence
$S_{p,1},S_{p,2}, \dots, S_{p,n_p}$, and a
good  initial guess for the coordinates $x_{p,i}'$.
Using the inverse of the equation (\ref{pere})
\beq
x_{p,i}''=S_{p,i}\sqrt{\frac{1-x_{p,i+1}'-x_{p,i-1}'}{a}},\;\; i=1,...,n_p
\eeq
we converge iteratively, at exponential rate, to the desired cycle points
$x_{p,i}$. Given the cycle points, the cycle stabilities and periods are
easily computed. The times and the stabilities of the short
periodic orbits for the H\'enon repeller \refeq{time_spread}
at $a=6$ are listed in table~\ref{gv:cycles}; in actual calculations we
use all prime cycles up to topological length $n=12$ (when needed,
all cycles up to length $n=20$ have been computed).
Once we have constructed a  table of lengths and stabilities,
extraction of the eigenvalues proceeds as  in the three-disk
example discussed above.

We consider first the discrete time approximation,
with the parameter $\alpha$ set equal to zero. Then
the lengths of the periodic orbits are simply $T_p=n_p$,
the spectrum in the complex
wave-number $k$ plane is  periodic  along
the real direction,  and it is sufficient to consider
the $0\leq \mbox{Re}(k) < 2\pi$ strip.  Fig.~\ref{gv1} shows
well separated eigenvalues for both the \cFd\ and
the \qFd, with no hint of a border of analyticity. In contrast,
a corresponding contour plot of \qS\ (not included here) shows very
clearly the finite radius of convergence, similar to what is
observed in figs.~\ref{qfselbrast}(a) and ~\ref{gv3}.

When we switch from a map to a model of a flow by
introducing a time spread, $\alpha\neq 0$
in (\ref{acti}), the spectrum is no longer periodic in
Re$(k)$.  In fig.~\ref{gv2} we plot the \cFd\ and
the \qFd\ resonances on the same scale as in fig.~\ref{gv1},
and in fig.~\ref{gv4} we plot a wider Re$(k)$ range.
Both \cFd\ and the \qFd\ converge as well as in the
$\alpha=0$ discrete time case
(where the theorem of sect.~\ref{ENTIRE} guarantees that at least the
\cFd\ is entire), and in the Re$(k)$ direction
a quasi-periodic pattern of eigenvalues can be observed,
similar to the 3-disk resonance spectrum of fig.~\ref{qf.cl.zeros}.

The unexpected feature of this spectrum is that the
leading quantum resonance in the $0\leq \mbox{Re}(k) < 2\pi$
domain has larger  imaginary part then the  leading resonances
in  the sectors $2\pi n\leq  \mbox{Re}(k) <2\pi(n+1)\;\; n=1,2,... $.
This is a consequence of the partial cancellation
of the contributions from the two fundamental cycles
$\bar{0}$ and $\bar{1}$ in the cycle expansion
(\ref{t_p_quant}),
due to relative minus sign arising from the Maslov phases.
A partial cancelation of the leading terms in the cycle
expansion diminishes the leading
coefficients in the $1/\zeta_0$ cycle expansion, hence the leading
zero has a larger imaginary part.
In the 3-disk example these contributions were positive
in the $A_1$  subspace, but in the $A_2$ subspace they partially
cancel in the same way as in our example.
For the classical \Fd, the weight (\ref{cl_weig})
depends only on the absolute magnitude of the stability, so
the  imaginary part of the leading zero of the \cFd\
in the domain $0\leq \mbox{Re}(k)<2\pi$  is smaller then the imaginary
part of the \qS's leading zero. This goes contrary to the
usual expectation that the quantum escape rate should be slower than the
classical one\rf{gasp}.

This model of a flow could also be applied to smooth potentials,
with other choices for the  Maslov indices,  and the other
forms of the action. We have computed the spectra with Maslov indices
appropriate to flows in smooth potentials,
and found qualitatively similar analytic properties of the \qFd.

\subsection{Baker's map}

As the third and the last illustration of the improved
convergence obtained by use of the \qFd s,
we evaluate both the \qFd\ and \qS\ for a model 2-$d$ map 
that takes the unit square into itself by the following
piecewise analytic transformations:
\\

\noindent
$\mbox{\ \ for \ \ } 0 < x_2 < 2/5$:
\bea
  f_1(x_1,x_2) &=& \frac{2}{5}  x_1 + x_2 (\frac{2}{5}-x_2)x_1(1-x_1)
        \continue
  f_2(x_1,x_2) &=& \frac{5}{2}  x_2 + 2 x_2 (\frac{2}{5}-x_2)x_1(1-x_1)
\nnu
\eea
\\ \noindent
$\mbox{\ \ for \ \ }  2/5 < x_2 < 1$:
\bea
  f_1(x_1,x_2) &=& \frac{3}{10}  x_1 + \frac{7}{10}
        \continue
  f_2(x_1,x_2) &=& \frac{5}{3}  (x_2- \frac{2}{5})
\, .
\label{12.1}
\eea
The model has a binary symbolic dynamics and no
particular physical motivation; it was introduced
in ref.~\cite{Rugh92} to illustrate numerically the theorem that the
\cFd\ for an Axiom A hyperbolic flow is entire. We study it here
merely as another
convenient model for numerical investigations of the convergence of the
contracting map equivalents of the \qFd\ and the \qS.
For this discrete time example, we assume no Maslov indices, and chose
the action to be equal to the (integer) time.
The ``\Gt'' for this map is defined by assigning cycles
weight $1/\sqrt{|\det(1-{\bf J}_p)|}$, with the corresponding ``\qS''.

The convergence of the cycle expansions (\ref{cyc_exp})
for the two determinants is illustrated by  fig.~\ref{hhr:deter}.
Starting with cycle length $n=13$, and up to $n=24$, the highest
cycle length computed, the \qFd\
performs better than the \qS.
The convergence is also illustrated
by the table~\ref{hhr:roots}, where the zeros
of the two determinants are shown.
Even though the map is not area preserving
and the ``\qS" is quite artificial, the convergence is much the
same as for the 3-disk repeller.

The numerical evidence from baker's map on whether the
\qFd\ and weighted \Fd s are entire is inconclusive; zeros of
$F_{-1}$,
$\Fqm$,
$F_{0}$,
$F_{1}$ cycle expansion truncations up to $n_{max}=22$ have qualitatively
the same distribution in the complex plane.
$1/\zeta(k)$ and $\Zqm(k)$
exhibit somewhat worse convergence than the \Fd s.
In all cases, only
a few leading zeros are true zeros; for example,
table~\ref{hhr:roots} lists 7 true zeros, the remaining 17  form a
``wedge" in the complex plane (see fig.~\ref{hhr:plot}),
which slowly drifts toward higher
Im$(k)$ with increased cycle truncation length.
\\

We conclude our discussion of numerics with a word of caution:
as we do not know how quickly the asymptotics should set in,
our numerical results can easily be misleading. For example, for a larger
disk-disk spacing, pre-asymptotic oscillations are visible in
fig.~2, and one might mistakenly conclude\rf{ER92} from such data that the
\cFd\ has a pole.
There is no substitute for
theorems that established that appropriate determinants are entire;
such oscillations make numerical convergence uncertain
already in the simplest 1-dimensional repellers\rf{Rosenqvist}.

\section{CONCLUSIONS AND SUMMARY}

In conclusion, we have tested numerically the conjecture
that the new approximation to the
quantum determinant, the \qFd, has better
analyticity properties than the commonly used \qS.
Existence of such determinant suggests a
starting approximation to the quantum propagator different from
the usual Van Vleck semi-classical propagator.
As we do not know an operator whose determinant is the \qFd, we have no
proof that such determinant is entire, only numerical evidence that its
convergence is superior to the \qS.
The new determinant could be of practical utility,
as for nice hyperbolic systems its convergence is superior to
that of the \qS s.

We were guided here by the Axiom A intuition developed by Smale, Ruelle,
and others:
if the dynamical evolution can be cast in terms of an evolution
operator multiplicative along the flow, if the corresponding mapping
(for ex., return map for a Poincar\'e section of the flow) is
analytic hyperbolic, and if the topology of the repeller is given by a finite
Markov partition, then the \Fd\  (\ref{cl_weig}) is entire.
An alternative approach, inspired by the theory of the
Riemann zeta function, is due to M.V.~Berry and J.~Keating\cite{BK90}.
The idea is to improve the periodic orbit expansions by imposing
unitarity as a  functional equation ansatz.
The cycle expansions used are the same as the
original ones\cite{AACI,CPR90}, but the philosophy is quite
different; the claim is that the optimal estimate for low eigenvalues of
classically chaotic quantum systems is obtained by taking the
real part of the cycle expansion of the semi-classical zeta function,
cut at the appropriate cycle period.

The real life challenge are generic dynamical flows, which fit
neither schematization.
Unfortunately we know of no
smooth potential which is both Axiom A, and has bound states.
Most systems of interest are {\em not} of the ``Axiom A" category;
they are neither purely hyperbolic nor do they have a simple
symbolic dynamics grammar.
The crucial
ingredient for nice analyticity properties of zeta functions is
existence of finite grammar (coupled with uniform hyperbolicity).
{}From hyperbolic dynamics point of view, the Riemann zeta is
perhaps the worst possible example; understanding the symbolic
dynamics would amount to being able to give a finite
grammar definition of all primes.
Hyperbolic dynamics suggests that
a generic ``chaotic" dynamical system should be approached by
a sequence of finite grammar approximations\cite{AACI}, pretty much
as a ``generic" number is approached by  a sequence of
continued fractions.

The dynamical systems that we are {\em really} interested in -
for example, smooth bound Hamiltonian potentials -
are presumably never really chaotic,
and it is still unclear what intuition is more rewarding:
are quantum spectra of chaotic dynamics in
smooth bound Hamiltonian potentials more like
zeros of Riemann zetas or zeros of \dzeta s?
\\

{\bf Note added in proof:}
Since completion of this work, Vattay {\em et al}\cite{CV93}
have succeeded in constructing a multiplicative evolution
operator for semi-classical quantum mechanics, with
the corresponding determinant
\beq
F(\beta,E)\,=\,\exp\left(-\sum_{p,r}
        {1 \over r}
      \frac{ |\Lambda_p|^{-r\beta}
             e^{i(S_p(k)/\hbar + \nu_p \pi/2)r} }
           {(1-1/\Lambda_p^r)^2 (1-1/\Lambda_p^{2r})}
\right)
\,.
\label{Vatt_det}
\eeq
This determinant is entire for the Axiom A flows. The results
presented in this paper remain valid, but the new formulation
makes it possible to reformulate our conjectures as
theorems. In particular,
the \qFd\ can now be written as a ratio of two Vattay determinants
\[ 
\Fqm(k) = { F({1 \over 2},k)
              \over
              F({5 \over 2},k)  }
\,,
\] 
and as explained in sect.~\ref{ABSCISSA},
the abcissa of convergence for such ratio is given by the
upper bound on the leading zero of $F({5 \over 2},k)$.
For example, for the 3-disk $R:a = 6:1$ system  we find
\[
           \mbox{Im}(k_{ac}) =  -  1.276625955 \dots \,
\]
so the \qFd\ can be continued almost a factor 2 down in the complex $k$ plane
beyond the region of applicability of the \qS. This explains why for all
practical purposes the \qFd s behave as entire functions: only careful
numerics reveals the difference between the  \qFd\ and the \Vd\ in higher
order terms in cycle expansions. The difference is illustrated by
fig.~\ref{vattay_d};
the suspicious straight section visible in the \qFd\,
($\triangle$) in fig.~\ref{fig2}
turns out to indeed indicate a pole,
while the \Vd\ converges super-exponentially.

\acknowledgements
G.V. is grateful to
the Sz\'echenyi Foundation and OTKA F4286 for the support, and
to the Center for Chaos and Turbulence Studies, Niels Bohr Institute, for
hospitality. P.C. thanks the Carlsberg Fundation for support.





\begin{figure}
\caption[]{
For a analytic hyperbolic map, specifying
the contracting coordinate $w_h$ at the initial rectangle and
the expanding coordinate $z_v$ at the image rectangle
defines a unique trajectory between the two rectangles.
In particular, $w_v$ and $z_h$ (not shown) are uniquely specified.
}
\label{hhr:rects}
\end{figure}

\begin{figure}
\caption[]{
$\log_{10} | C_n |$, the contribution of cycles of
topological length $n$
to the cycle expansion $\sum C_n z^n$ for 3-disk repeller.
Shown are:
($\circ$) $1/\zeta_0$,
($\nabla$) the \qS,
($\Box$) $1/\zeta_0 \zeta_1^2$,
and
($\triangle$) the \qFd.
Exponential falloff implies that $1/\zeta_0$ and the \qS\
have the same leading pole, cancelled in the $1/\zeta_0 \zeta_1^2$
product. For comparison, ($\Diamond$) the \cFd\ coefficients are plotted
as well; cycle expansions for both \Fd s appear
to follow the asymptotic estimate $C_n \approx \Lambda^{-n^{3/2}}$.
$A_1$ symmetric subspace,
with center spacing - disk radius ratio $R:a= 3:1$,
evaluated at the lowest resonance, wave number
$k=7.8727 - 0.3847\,i $, maximal cycle length $n=8$.
}
\label{fig1}
\end{figure}

\begin{figure}
\caption[]{ Same as fig.~\ref{fig1}, but with $ R : a = 6 : 1 $.
This illustrates possible pitfalls of numerical tests of asymptotics;
the \qFd\  appears to have the same pole as
the quantum $1/\zeta_0 \zeta_1^2$, but
as we have no estimate on the size of pre-asymptotic oscillations in
cycle expansions, it is difficult to draw reliable conclusions from
such numerics.}
\label{fig2}
\end{figure}

\begin{figure}
\caption[]{Leading resonances in the  3-disk repeller $A_1$ subspace,
(a) for the \cFd, and (b)
the 952 leading resonances of the \qFd\ $\Fqm$.
Ratio $R:a = 6:0$, cycle expansions truncated at cycle length $n=8$.
}
\label{qf.cl.zeros}
\end{figure}

\begin{figure}
\caption[]{ ``Difference" between the classical $F_0$ and $F_1$
spectrum for the 3-disk repeller.
All resonances $s_{0,\alpha}$ of $F_0$ that fall within
$|\mbox{Re}( s_{0,\alpha}) -\mbox{Re}(s_{1,\beta})| < 0.1$,
$|\mbox{Im}( s_{0,\alpha}) -\mbox{Im}(s_{1,\beta})| < 0.08$,
of a resonance $s_{1,\beta}$ of $F_1$,  are deleted from
fig.~\ref{qf.cl.zeros}. Note that while the non-leading families
of resonances $s_{0,\alpha}$ almost coincide with the
leading  $s_{1,\beta}$ resonances, they are
not degenerate. With the resonances associated with the $F_1$
spectrum removed, families that connect nonleading resonances
at Re$(s)=0$ with the leading part of the resonance spectrum for
larger Re$(s)$ are clearly visible.
$R:a = 6:1$, $A_1$ subspace, maximal cycle length $n=8$.   }
\label{F_0-F_1}
\end{figure}

\begin{figure}
\caption[]{Complex $s$ plane countor plot comparison of (a) the \qS\
$\log|\Zqm(s)|$ with (b) the \qFd\ $\log|\Fqm(s)|$. The border of
the convergence of the \qS\ agrees with the location of the
abscissa of absolute convergence,
given by the $\hat{F}_{1/2}$ leading eigenvalue at
$\mbox{Re}(s) = 0$,
$\mbox{Im}(s) = -0.699110157151\dots$.
The \qFd\ can be continued at
least a factor 3 further down in the complex plane.
3-disk repeller, $R:a = 6:1$, $A_1$ subspace, maximal cycle length $n=8$.}
\label{qfselbrast}
\end{figure}

\begin{figure}
\caption[]{Contour plot of $\log_{10} | F(s) |$ for
({a})
the \cFd,  and for
({b})
the \qFd\ for the H\'enon map, $\alpha=0$ in \refeq{time_spread},
prime cycles up to topological length $12$.}
\label{gv1}
\end{figure}

\begin{figure}
\caption[]{Contour plot of $\log_{10} | F(s) |$ for
({a})
the \cFd,  and for
({b})
the \qFd\ for the model flow \refeq{time_spread} with $\alpha=1/2$,
prime cycles up to topological length $12$.}
\label{gv2}
\end{figure}

\begin{figure}
\caption[]{
Contour plot of $\log_{10} | F(s) |$ for
the \qFd; same as fig.~\ref{gv2}, but for a wider range of $Re(k)$.
           }
\label{gv4}
\end{figure}

\begin{figure}
\caption[]{Same as fig.~\ref{gv2} for the \qS. The single leading
pole of the \qS\ manifests itself as the border of the convergence;
only two leading eigenvalues from fig.~\ref{gv2}({b}) remain within the
domain of convergence.}
\label{gv3}
\end{figure}

\begin{figure}
\caption[]{$\log_{10} | C_n |$, the contribution of cycles of
topological length $n$ to the cycle expansion $\sum C_n z^n$, for
the ($\Box$) ``\qS" and ($\circ$) ``\qFd" for the baker's map \refeq{12.1}.}
\label{hhr:deter}
\end{figure}

\begin{figure}
\caption[]{
The ``\qFd" zeros for the
baker's map \refeq{12.1}
computed from all cycles to cycle length $n_{max}=22$.
A few leading zeros are true zeros;
table~\ref{hhr:roots} lists the 7 true zeros, the remaining 17  forming a
``wedge" in the complex plane which slowly drifts toward higher
Im$(k)$ with increased cycle truncation length.
          }
\label{hhr:plot}
\end{figure}

\begin{figure}
\caption[]{Same parameter values 3-disk system as fig.~\ref{fig1}:
($\circ$) the \qFd\ compared with ($\Box$) the \Vd.
While the \qFd\  is expected to have a pole at
Im$(k)=-1.276625955 \dots$,
the \Vd\ should be entire, and exhibits numerically
faster than exponential convergence.
}
\label{vattay_d}
\end{figure}


\onecolumn
\mediumtext

\narrowtext
\begin{table}
\caption[]{ All periodic orbits up to 6 bounces
for the Hamiltonian H\'enon mapping \refeq{time_spread} with $a=6$.
Listed are the topological length of the cycle,
its expanding eigenvalue $\Lambda_p$,
the variation in the period of the cycle, and its binary code.
\\
  }
\begin{tabular}{|c|r|r|r|}
\tableline
$n_p$ & $\Lambda_p$~~~~~~~~~~~~
                            &    $\sum x_{p,i}$~~~~~~~~~
                                                 & code\\ \hline
1 & 0.71516752438$\times 10^1$ & -0.6076252185107 & 0 \\
1 &-0.29528463259$\times 10^1$ &  0.2742918851774 & 1 \\
   \hline
2 &-0.98989794855$\times 10^1$ &  0.3333333333333 & 10 \\
   \hline
3 &-0.13190727397$\times 10^3$ & -0.2060113295833 & 100 \\
3 & 0.55896964996$\times 10^2$ &  0.5393446629166 & 110 \\
   \hline
4 &-0.10443010730$\times 10^4$ & -0.8164965809277 & 1000 \\
4 & 0.57799826989$\times 10^4$ &  0.0000000000000 & 1100 \\
4 &-0.10368832509$\times 10^3$ &  0.8164965809277 & 1110 \\
   \hline
5 &-0.76065343718$\times 10^4$ & -1.4260322065792 & 10000 \\
5 & 0.44455240007$\times 10^4$ & -0.6066540777738 & 11000 \\
5 & 0.77020248597$\times 10^3$ &  0.1513755016405 & 10100 \\
5 &-0.71068835616$\times 10^3$ &  0.2484632276044 & 11100 \\
5 &-0.58949885284$\times 10^3$ &  0.8706954728949 & 11010 \\
5 & 0.39099424812$\times 10^3$ &  1.0954854155465 & 11110 \\
   \hline
6 &-0.54574527060$\times 10^5$ & -2.0341342556665 & 100000 \\
6 & 0.32222060985$\times 10^5$ & -1.2152504370215 & 110000 \\
6 & 0.51376165109$\times 10^4$ & -0.4506624359329 & 101000 \\
6 &-0.47846146631$\times 10^4$ & -0.3660254037844 & 111000 \\
6 &-0.63939998436$\times 10^4$ &  0.3333333333333 & 110100 \\
6 &-0.63939998436$\times 10^4$ &  0.3333333333333 & 101100 \\
6 & 0.39019387269$\times 10^4$ &  0.5485837703548 & 111100 \\
6 & 0.10949094597$\times 10^4$ &  1.1514633582661 & 111010 \\
6 &-0.10433841694$\times 10^4$ &  1.3660254037844 & 111110 \\
   \hline
\end{tabular}
\label{gv:cycles}
\end{table}
\bigskip

\narrowtext
\begin{table}
\caption[]{The ``quantum" eigenvalues obtained from the \qS\
$\Zqm(s)$ and the \qFd\ $\Fqm(s)$ for the baker's map.
Only the leading eigenvalues are expected to coincide.
The digits listed correspond to those unchanged between the
cycle length $n=23$ and $n=24$ truncations.
\\
          }
\begin{tabular}{|c| l | ll |}
\hline
    & \multicolumn{1}{c|}{$\Zqm(s)$} &
      \multicolumn{2}{c|}{$\Fqm(s)$}
      \\
k   &  \multicolumn{1}{c|} { Real part}
    &  \multicolumn{1}{c} { Real part }
    &  \multicolumn{1}{c|}{ Im. part }
    \\ \hline

1   &     0.34051779186
    &    +0.3405177918632516      &    \\
2   &     -0.39
    &    -0.3500689652275 &    \\
3   &
    &    -0.7175716624      &    \\
4-5 &
    &    -1.099684                   &  $\pm$ 0.115118    \\
6-7 &
    &    -1.59                       &  $\pm$ 0.227    \\
      \hline
\end{tabular}
\label{hhr:roots}
\end{table}

\end{document}